\documentclass[10pt,a4paper,onecolumn]{article}
\usepackage{marginnote}
\usepackage{graphicx}
\usepackage{xcolor}
\usepackage{authblk,etoolbox}
\usepackage{titlesec}
\usepackage{calc}
\usepackage{tikz}
\usepackage{hyperref}
\hypersetup{colorlinks,breaklinks=true,
            urlcolor=[rgb]{0.0, 0.5, 1.0},
            linkcolor=[rgb]{0.0, 0.5, 1.0}}
\usepackage{caption}
\usepackage{tcolorbox}
\usepackage{amssymb,amsmath}
\usepackage{ifxetex,ifluatex}
\usepackage{seqsplit}
\usepackage{xstring}

\usepackage{float}
\let\origfigure\figure
\let\endorigfigure\endfigure
\renewenvironment{figure}[1][2] {
    \expandafter\origfigure\expandafter[H]
} {
    \endorigfigure
}

\usepackage{fixltx2e} 
\usepackage[
  backend=biber,
]{biblatex}
\bibliography{paper.bib}

\let\textttOrig=\texttt
\def\texttt#1{\expandafter\textttOrig{\seqsplit{#1}}}
\renewcommand{\seqinsert}{\ifmmode
  \allowbreak
  \else\penalty6000\hspace{0pt plus 0.02em}\fi}

\makeatletter
\let\href@Orig=\href
\def\href@Urllike#1#2{\href@Orig{#1}{\begingroup
    \def\Url@String{#2}\Url@FormatString
    \endgroup}}
\def\href@Notdoi#1#2{\def\tempa{#1}\def\tempb{#2}%
  \ifx\tempa\tempb\relax\href@Urllike{#1}{#2}\else
  \href@Orig{#1}{#2}\fi}
\def\href#1#2{%
  \IfBeginWith{#1}{https://doi.org}%
  {\href@Urllike{#1}{#2}}{\href@Notdoi{#1}{#2}}}
\makeatother

\newlength{\cslhangindent}
\newlength{\csllabelwidth}
\newenvironment{CSLReferences}[3] 
 {
  \setlength{\parindent}{0pt}
  \ifodd #1 \everypar{\setlength{\hangindent}{\cslhangindent}}\ignorespaces\fi
  \ifnum #2 > 0
  \setlength{\parskip}{#2\baselineskip}
  \fi
 }%
 {}
\usepackage{calc}

\usepackage[top=3.5cm, bottom=3cm, right=1.5cm, left=1.0cm,
            headheight=2.2cm, reversemp, includemp, marginparwidth=4.5cm]{geometry}

\titleformat{\section}
  {\normalfont\sffamily\Large\bfseries}
  {}{0pt}{}
\titleformat{\subsection}
  {\normalfont\sffamily\large\bfseries}
  {}{0pt}{}
\titleformat{\subsubsection}
  {\normalfont\sffamily\bfseries}
  {}{0pt}{}
\titleformat*{\paragraph}
  {\sffamily\normalsize}

\usepackage{fancyhdr}
\makeatletter
\let\ps@plain\ps@fancy
\fancyheadoffset[L]{4.5cm}
\fancyfootoffset[L]{4.5cm}

\definecolor{linky}{rgb}{0.0, 0.5, 1.0}

\newtcolorbox{repobox}
   {colback=red, colframe=red!75!black,
     boxrule=0.5pt, arc=2pt, left=6pt, right=6pt, top=3pt, bottom=3pt}

\newcommand{\ExternalLink}{%
   \tikz[x=1.2ex, y=1.2ex, baseline=-0.05ex]{%
       \begin{scope}[x=1ex, y=1ex]
           \clip (-0.1,-0.1)
               --++ (-0, 1.2)
               --++ (0.6, 0)
               --++ (0, -0.6)
               --++ (0.6, 0)
               --++ (0, -1);
           \path[draw,
               line width = 0.5,
               rounded corners=0.5]
               (0,0) rectangle (1,1);
       \end{scope}
       \path[draw, line width = 0.5] (0.5, 0.5)
           -- (1, 1);
       \path[draw, line width = 0.5] (0.6, 1)
           -- (1, 1) -- (1, 0.6);
       }
   }

\patchcmd{\@maketitle}{center}{flushleft}{}{}
\patchcmd{\@maketitle}{center}{flushleft}{}{}
\patchcmd{\@maketitle}{\LARGE}{\LARGE\sffamily}{}{}
\def\maketitle{{%
  
  \AB@maketitle}}
\makeatletter
\renewcommand\AB@affilsepx{ \protect\Affilfont}
\renewcommand\AB@affilnote[1]{{\bfseries #1}\hspace{3pt}}
\renewcommand{\affil}[2][]%
   {\newaffiltrue\let\AB@blk@and\AB@pand
      \if\relax#1\relax\def\AB@note{\AB@thenote}\else\def\AB@note{#1}%
        \setcounter{Maxaffil}{0}\fi
        \begingroup
        \let\href=\href@Orig
        \let\texttt=\textttOrig
        \let\protect\@unexpandable@protect
        \def\thanks{\protect\thanks}\def\footnote{\protect\footnote}%
        \@temptokena=\expandafter{\AB@authors}%
        {\def\\{\protect\\\protect\Affilfont}\xdef\AB@temp{#2}}%
         \xdef\AB@authors{\the\@temptokena\AB@las\AB@au@str
         \protect\\[\affilsep]\protect\Affilfont\AB@temp}%
         \gdef\AB@las{}\gdef\AB@au@str{}%
        {\def\\{, \ignorespaces}\xdef\AB@temp{#2}}%
        \@temptokena=\expandafter{\AB@affillist}%
        \xdef\AB@affillist{\the\@temptokena \AB@affilsep
          \AB@affilnote{\AB@note}\protect\Affilfont\AB@temp}%
      \endgroup
       \let\AB@affilsep\AB@affilsepx
}
\makeatother

\renewcommand\Affilfont{\sffamily\small\mdseries}
\ifnum 0\ifxetex 1\fi\ifluatex 1\fi=0 
  \usepackage[T1]{fontenc}
  \usepackage[utf8]{inputenc}

\else 
  \ifxetex
    \usepackage{mathspec}
    \usepackage{fontspec}

  \else
    \usepackage{fontspec}
  \fi
  \defaultfontfeatures{Ligatures=TeX,Scale=MatchLowercase}

\fi
\IfFileExists{upquote.sty}{\usepackage{upquote}}{}
\IfFileExists{microtype.sty}{%
\usepackage{microtype}
\UseMicrotypeSet[protrusion]{basicmath} 
}{}

\usepackage{hyperref}
\hypersetup{unicode=true,
            pdftitle={Sunny.jl: A Julia Package for Spin Dynamics},
            pdfborder={0 0 0},
            breaklinks=true}
\let\addcontentslineOrig=\addcontentsline
\def\addcontentsline#1#2#3{\bgroup
  \let\texttt=\textttOrig\addcontentslineOrig{#1}{#2}{#3}\egroup}
\let\markbothOrig\markboth
\def\markboth#1#2{\bgroup
  \let\texttt=\textttOrig\markbothOrig{#1}{#2}\egroup}
\let\markrightOrig\markright
\def\markright#1{\bgroup
  \let\texttt=\textttOrig\markrightOrig{#1}\egroup}

\usepackage{graphicx,grffile}
\makeatletter
\def\maxwidth{\ifdim\Gin@nat@width>\linewidth\linewidth\else\Gin@nat@width\fi}
\def\maxheight{\ifdim\Gin@nat@height>\textheight\textheight\else\Gin@nat@height\fi}
\makeatother
\setkeys{Gin}{width=\maxwidth,height=\maxheight,keepaspectratio}
\IfFileExists{parskip.sty}{%
\usepackage{parskip}
}{
\setlength{\parindent}{0pt}
\setlength{\parskip}{6pt plus 2pt minus 1pt}
}
\providecommand{\tightlist}{%
  \setlength{\itemsep}{0pt}\setlength{\parskip}{0pt}}
\ifx\paragraph\undefined\else
\let\oldparagraph\paragraph
\renewcommand{\paragraph}[1]{\oldparagraph{#1}\mbox{}}
\fi
\ifx\subparagraph\undefined\else
\let\oldsubparagraph\subparagraph
\renewcommand{\subparagraph}[1]{\oldsubparagraph{#1}\mbox{}}
\fi

\title{Sunny.jl: A Julia Package for Spin Dynamics\footnote{This
  manuscript has been authored by UT-Battelle, LLC, under contract
  DE-AC05-00OR22725 with the US Department of Energy (DOE). The US
  government retains and the publisher, by accepting the article for
  publication, acknowledges that the US government retains a
  nonexclusive, paid-up, irrevocable, worldwide license to publish or
  reproduce the published form of this manuscript, or allow others to do
  so, for US government purposes. DOE will provide public access to
  these results of federally sponsored research in accordance with the
  DOE Public Access Plan
  (https://www.energy.gov/doe-public-access-plan).}}

        \author[1, 2]{David Dahlbom}
          \author[3]{Hao Zhang}
          \author[4]{Cole Miles}
          \author[5, 6]{Sam Quinn}
          \author[7]{Alin Niraula}
          \author[7]{Bhushan Thipe}
          \author[8]{Matthew Wilson}
          \author[3]{Sakib Matin}
          \author[9]{Het Mankad}
          \author[9]{Steven Hahn}
          \author[1]{Daniel Pajerowski}
          \author[2, 10]{Steve Johnston}
          \author[11]{Zhentao Wang}
          \author[12, 13, 14]{Harry Lane}
          \author[15]{Ying Wai Li}
          \author[7]{Xiaojian Bai}
          \author[5]{Martin Mourigal}
          \author[2]{Cristian D. Batista}
          \author[3]{Kipton Barros}
    
      \affil[1]{Neutron Scattering Division, Oak Ridge National
Laboratory}
      \affil[2]{Department of Physics and Astronomy, University of
Tennessee}
      \affil[3]{Theoretical Division and CNLS, Los Alamos National
Laboratory}
      \affil[4]{Kodiak Robotics}
      \affil[5]{School of Physics, Georgia Institute of Technology}
      \affil[6]{Department of Physics and Astronomy, Univeriy of
California, Los Angeles}
      \affil[7]{Department of Physics and Astronomy, Louisiana State
University}
      \affil[8]{X-Computational Physics Division, Los Alamos National
Laboratory}
      \affil[9]{Computer Science and Mathematics Division, Oak Ridge
National Laboratory}
      \affil[10]{Institute for Advanced Materials and Manufacturing,
Unversity of Tennessee}
      \affil[11]{Center for Correlated Matter and School of Physics,
Zhejiang University}
      \affil[12]{Department of Physics and Astronomy, University of
Manchester}
      \affil[13]{The University of Manchester at Harwell, University of
Manchester}
      \affil[14]{School of Physics and Astronomy, University of St
Andrews}
      \affil[15]{Computer, Computational, and Statistical Sciences
Division, Los Alamos National Laboratory}
  \date{\vspace{-7ex}}

\begin{document}
\maketitle

\marginpar{

  \begin{flushleft}
  \sffamily\small

  {\bfseries DOI:} \href{https://doi.org/DOI unavailable}{\color{linky}{DOI unavailable}}

  \vspace{2mm}

  {\bfseries Software}
  \begin{itemize}
    \setlength\itemsep{0em}
    \item \href{N/A}{\color{linky}{Review}} \ExternalLink
    \item \href{NO_REPOSITORY}{\color{linky}{Repository}} \ExternalLink
    \item \href{DOI unavailable}{\color{linky}{Archive}} \ExternalLink
  \end{itemize}

  \vspace{2mm}

  \par\noindent\hrulefill\par

  \vspace{2mm}

  {\bfseries Editor:} \href{https://example.com}{Pending
Editor} \ExternalLink \\
  \vspace{1mm}
    {\bfseries Reviewers:}
  \begin{itemize}
  \setlength\itemsep{0em}
    \item \href{https://github.com/Pending Reviewers}{@Pending
Reviewers}
    \end{itemize}
    \vspace{2mm}

  {\bfseries Submitted:} N/A\\
  {\bfseries Published:} N/A

  \vspace{2mm}
  {\bfseries License}\\
  Authors of papers retain copyright and release the work under a Creative Commons Attribution 4.0 International License (\href{http://creativecommons.org/licenses/by/4.0/}{\color{linky}{CC BY 4.0}}).

  \end{flushleft}
}

\hypertarget{summary}{%
\section{Summary}\label{summary}}

Sunny is a Julia package designed to serve the needs of the quantum
magnetism community. It supports the specification of a very broad class
of spin models and a diverse suite of numerical solvers. These include
powerful methods for simulating spin dynamics both in and out of
equilibrium. Uniquely, it features a broad generalization of classical
and semiclassical approaches to SU(\emph{N}) coherent states, which is
useful for studying systems exhibiting strong spin-orbit coupling or
local entanglement effects. Sunny also offers a well-developed framework
for calculating the dynamical spin structure factor, enabling direct
comparison with scattering experiments. Ease of use is a priority, with
tools for symmetry-guided modeling and interactive visualization.

\hypertarget{statement-of-need}{%
\section{Statement of need}\label{statement-of-need}}

Progress in quantum magnetism depends on the development of accurate
models of magnetic materials. Scattering techniques, such inelastic
neutron scattering (INS) and resonant inelastic X-ray scattering (RIXS),
are among the most informative methods available for probing the
dynamics of quantum magnets, yielding the dynamical spin structure
factor \(\mathcal{S}(\mathbf{q},\omega)\) as experimental output. To
evaluate the validity of a hypothetical model, it is necessary to
calculate \(\mathcal{S}(\mathbf{q}, \omega)\) theoretically. This is
generally an intractable problem that must be treated numerically or
with various approximation schemes. The difficulty of this step
represents a bottleneck in the development of accurate models and
impedes the advancement of our understanding of quantum materials.

The Sunny project is a collaborative effort among theorists,
experimentalists and computational scientists aimed at developing
theoretical and numerical methodologies for modeling realistic quantum
magnets. The central product of this effort is the Sunny software
package, which makes recent theoretical advances available in a form
readily accessible to students and researchers. Distinguishing features
of Sunny include:

\begin{itemize}
\tightlist
\item
  Symmetry analysis tools that facilitate model specification,
  visualization and data retrieval.
\item
  A suite of optimizers, Monte Carlo samplers, and spin dynamics solvers
  that can all be applied to the same system specification.
\item
  Implementation of the SU(\emph{N}) coherent state formalism for
  classical and semiclassical calculations.
\item
  An interface tailored toward the needs of scattering scientists, with
  tools for integrating scattering intensities over regions of
  reciprocal space.
\item
  Code written entirely in Julia, a language that can achieve speeds
  comparable to C++ or Fortran while offering an interactive workflow
  that will be familiar to users of Python and Matlab.
\item
  A well documented codebase, an extensive collection of correctness
  tests, a website featuring many tutorials, and an active Slack channel
  where users can ask questions.
\end{itemize}

There are a number of existing codes that can calculate
\(\mathcal{S}(\mathbf{q},\omega)\) using linear spin wave theory (LSWT),
some of which have served as inspiration to the Sunny project {[}Rotter
(2004); (2024a); Petit \& Damay (2016); weber:2016; (2024b){]}. The
symmetry analysis tools of SpinW in particular have served as a model
(Toth \& Lake, 2015). There are also codes that perform classical spin
simulations using Landau-Lifshitz (LL) dynamics (2024c; Evans et al.,
2014; Müller et al., 2019). Sunny is unique in offering both approaches
and generalizing them through a formalism based on SU(\emph{N}) coherent
states (Muniz et al., 2014; H. Zhang \& Batista, 2021). Sunny
additionally permits completely general single-ion anisotropies and
coupling of multipolar moments; provides an efficient implementation of
long-range dipole-dipole interactions; automates the application of a
number of quantum renormalizations (Dahlbom et al., 2023); and offers
iterative solvers for efficient LSWT on large magnetic cells (Lane et
al., 2024).

The value of collecting all these tools together in a modern,
easy-to-use package is evidenced by the large number of publications
that have already made use of Sunny, a partial list of which is
maintained on the GitHub wiki (2024d). We note a number of experimental
studies that have relied on Sunny for analysis (Bai et al., 2021, 2023;
Do et al., 2023; Kim et al., 2023; Lee et al., 2023; Na et al., 2024;
Nagl et al., 2024; Paddison et al., 2024; Park et al., 2023; Park, Sala,
et al., 2024; Park, Ghioldi, et al., 2024; Park, Cho, et al., 2024;
Scheie et al., 2023); as well as theoretical and methodological works
(Dahlbom, Brooks, et al., 2024; Dahlbom, Thomas, et al., 2024; H. Zhang
et al., 2023; Hao Zhang \& Lin, 2024). Additional papers documenting the
theoretical and algorithmic advances that have enabled the development
of Sunny are discussed below.

\hypertarget{feature-overview}{%
\section{Feature Overview}\label{feature-overview}}

\hypertarget{symmetry-analysis}{%
\subsection{Symmetry analysis}\label{symmetry-analysis}}

By unifying and extending existing open source frameworks for the
symmetry analysis of crystals -- including Spglib (Togo et al., 2024),
Brillouin.jl (2024e), and CrystalInfoFramework.jl (2024f) -- Sunny
facilitates the process of determining the complete set of interactions
allowed by spacegroup symmetries. Similarly, any interaction specified
on a site or bond will be automatically propagated to all
symmetry-equivalent sites and bonds, as required by the spacegroup
symmetries. Models may also be specified according to symmetry
properties and subsequently made ``inhomogenous,'' allowing the
arbitrary modification of pair interactions and site properties without
regard to symmetry constraints. This greatly facilitates the modeling of
systems exhibiting chemical disorder. Finally, the symmetry information
enables convenient specification of paths and slices through reciprocal
space, aiding visualization and comparison to experimental data. All
these tools can be applied just as easily to a user-specified crystal or
to a crystal loaded from an industry-standard CIF file (Hall et al.,
1991).

\begin{figure}
\centering
\includegraphics{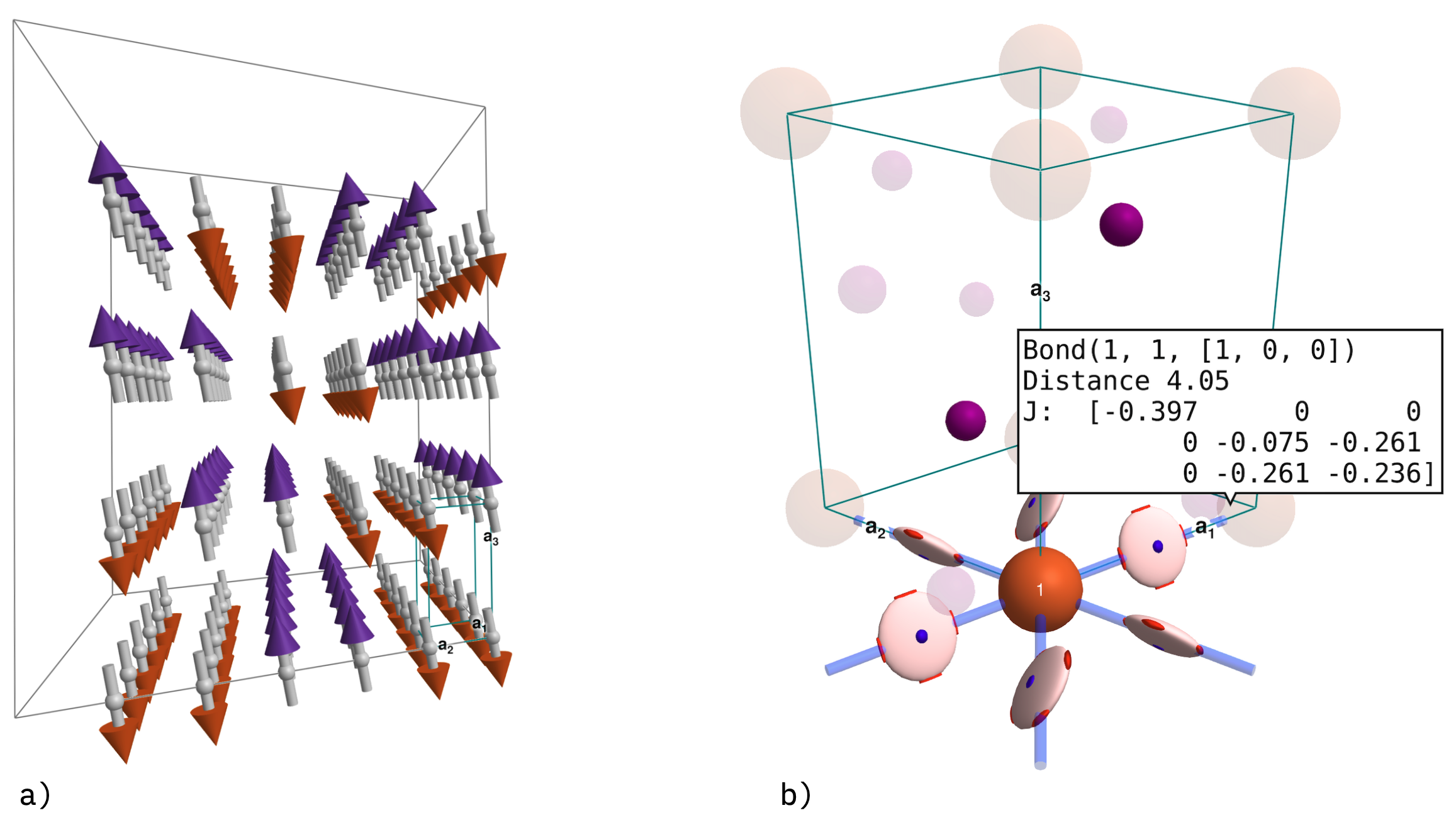}
\caption{a) Ground state of \(\mathrm{FeI}_{2}\), found using Sunny's
\texttt{minimize\_energy!} function and visualized with
\texttt{plot\_spins}. b) The crystal of \(\mathrm{FeI}_2\) visualized
with the \texttt{view\_crystal} function. Hovering the cursor over a
bond reveals the exchange interaction, if already assigned, or a general
expression for all symmetry-allowed interactions. \label{fig:symmetry}}
\end{figure}

\hypertarget{visualization}{%
\subsection{Visualization}\label{visualization}}

\begin{figure}
\centering
\includegraphics{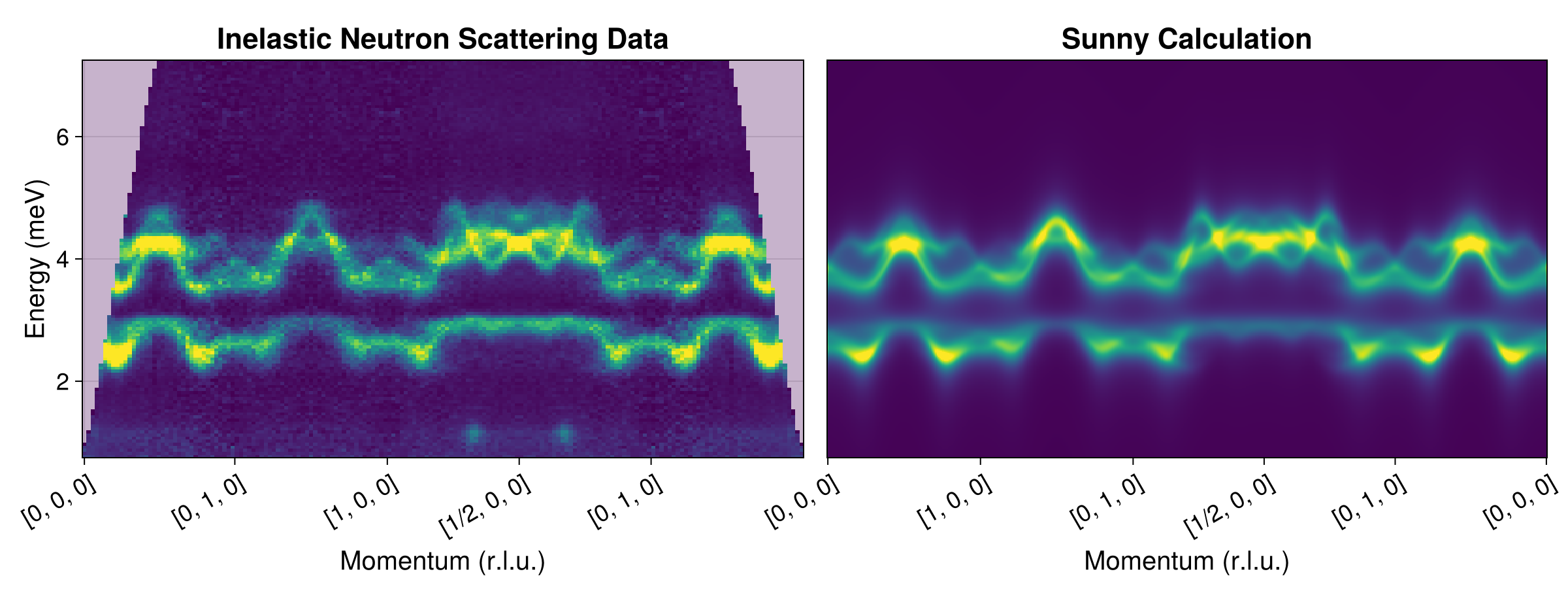}
\caption{\emph{Left}: Scattering intensities of \(\mathrm{FeI}_2\) as
measured on the SEQUOIA instrument at the Spallation Neutron Source, Oak
Ridge National Laboratory (Bai et al., 2021). \emph{Right}: Predicted
scattering intensities calculated with Sunny's SU(\emph{N}) linear spin
wave solver. The figure was generated with Sunny's data retrieval and
plotting functions. \label{fig:Sqw}}
\end{figure}

Both the symmetry analysis and data retrieval features of Sunny include
3D visualization tools built on the Makie package (Danisch \&
Krumbiegel, 2021). These can be used to plot spin configurations,
investigate the symmetries of a crystal \autoref{fig:symmetry}, generate
animations of dynamic behavior, and plot the predicted results of
scattering experiments \autoref{fig:Sqw}.

\hypertarget{sun-formalism-and-system-modes}{%
\subsection{\texorpdfstring{SU(\emph{N}) Formalism and System
Modes}{SU(N) Formalism and System Modes}}\label{sun-formalism-and-system-modes}}

Traditional classical and semiclassical approaches to spin dynamics are
based on the assignment of a classical dipole to each lattice site.
Recent theoretical work has generalized this picture, replacing dipoles
with richer objects, namely SU(\emph{N}) coherent states. Such states
capture the full structure of an \emph{N}-level quantum system. Setting
\(N = 2s + 1\) enables the faithful representation of a quantum
spin-\(s\) and of the crystal field levels of a single-ion. The
formalism can also be adapted to model local entanglement effects, where
this entanglement may be between the spin and orbital degrees of freedom
on a single site or within a cluster of spins on different sites.

The SU(\emph{N}) formalism applies equally to LSWT calculations (Muniz
et al., 2014) and classical spin dynamics (H. Zhang \& Batista, 2021).
Users can access this framework simply by setting the ``mode'' of a spin
system to \texttt{:SUN}. Sunny also offers a \texttt{:dipole} mode,
which is similar to the traditional classical approach but includes
quantum renormalizations of biquadratic and single-ion anisotropy terms
(Dahlbom et al., 2023). Finally, there is a mode that implements the
traditional approach without any additional corrections,
\texttt{:dipole\_uncorrected}. Most Sunny features are supported in all
modes.

\hypertarget{optimization-and-monte-carlo-tools}{%
\subsection{Optimization and Monte Carlo
Tools}\label{optimization-and-monte-carlo-tools}}

Identifying a classical ground state is often the first step when
calculating the scattering response of a magnet. Sunny provides several
tools for finding such states, including gradient optimizers built on
the Optim.jl package (Mogensen \& Riseth, 2018). These optimizers work
on supercells as well as spiral orderings. Sophisticated Monte Carlo
tools are also provided, which can be used both to anneal into ground
states and to estimate finite temperature statistics. In particular, the
classical dynamics can be run with Langevin coupling to a thermal bath
(Dahlbom, Miles, et al., 2022), and samplers are provided that implement
the Wang-Landau (Wang \& Landau, 2001) and parallel tempering algorithms
(Swendsen \& Wang, 1986).

\hypertarget{linear-spin-wave-theory-lswt}{%
\subsection{Linear Spin Wave Theory
(LSWT)}\label{linear-spin-wave-theory-lswt}}

Sunny has extensive support for LSWT calculations, including for systems
with arbitrarily complex single-ion anisotropies, general bilinear and
biquadratic interactions, and long-range dipole-dipole interactions.
Like SpinW, Sunny provides efficient LSWT calculations for systems that
exhibit incommensurate spiral orderings (Toth \& Lake, 2015).
Additionally, Sunny provides tools to efficiently calculate
\(\mathcal{S}(\mathbf{q},\omega)\) on very large magnetic cells using
iterative matrix-vector multiplications (Lane et al., 2024). The
simulation of large supercells is essential to study systems with
chemical disorder or complex magnetic orderings.

\hypertarget{classical-dynamics}{%
\subsection{Classical Dynamics}\label{classical-dynamics}}

The efficiency of LSWT calculations makes it the preferred tool when
studying magnets near zero temperature. At elevated temperatures or in
out-of-equilibrium conditions, however, classical dynamics becomes a
valuable technique. Sunny supports both traditional Landau-Lifshitz
dynamics and its generalization to SU(\emph{N}) coherent states (H.
Zhang \& Batista, 2021). Dissipationless trajectories are calculated
using a symplectic integration scheme (Dahlbom, Zhang, et al., 2022),
and a generalization of the stochastic Landau-Lifshitz-Gilbert equations
to SU(\emph{N}) coherent states (Dahlbom, Miles, et al., 2022) enables
the simulation of dynamics coupled to a thermal bath. This is
particularly valuable for simulating, e.g., thermal transport,
pump-probe experiments, and spin-glass relaxation.

\hypertarget{sunny-as-a-platform-for-future-developments}{%
\subsection{Sunny as a Platform for Future
Developments}\label{sunny-as-a-platform-for-future-developments}}

To make these existing features more widely available, work at ORNL is
underway to integrate Sunny into the Calvera platform for neutron data
analysis (Watson et al., 2022). Sunny itself can serve as a platform for
new solvers and analysis techniques, building on its mature model
specification and data retrieval features. Current efforts are directed
at supporting: the self-consistent Gaussian approximation for diffuse
scattering, enabling functionality inspired by (Paddison et al., 2024);
the modeling of local entanglement effects generated by spin-orbit
coupling or strongly coupled clusters of spins; non-perturbative
corrections to LSWT for the modeling of continua and bound states, which
can be probed in INS and terahertz spectroscopy experiments (Bai et al.,
2023; Legros et al., 2021); and observables relevant to RIXS
experiments.

\hypertarget{acknowledgements}{%
\section{Acknowledgements}\label{acknowledgements}}

We thank Mosé Giordano and Simon Danisch for valuable discussions. This
work was supported by the U.S. Department of Energy, Office of Science,
Office of Basic Energy Sciences, under Award Numbers DE-SC0022311,
DE-SC-0018660, and DE-SC0025426. The work was also sponsored by the
Laboratory Directed Research and Development Programs (LDRD) at Los
Alamos National Laboratory, managed by Triad National Security, LLC, and
at Oak Ridge National Laboratory, managed by UT-Battelle, LLC, for the
U. S. Department of Energy. Applications to inverse scattering were
partially supported by the National Science Foundation Materials
Research Science and Engineering Center program through the UT Knoxville
Center for Advanced Materials and Manufacturing (DMR-2309083). Z.W.
acknowledges support from the National Key Research and Development
Program of China (Grant No.~2024YFA1408303) and the National Natural
Science Foundation of China (Grant No.~12374124). H.L. acknowledges
funding from the Royal Commission for the Exhibition of 1851. The data
shown in Figure 2 was collected at the the Spallation Neutron Source, a
DOE Office of Science User Facility operated by the Oak Ridge National
Laboratory. Beam time was allocated to the SEQUOIA instrument on
proposal number IPTS-21166.

\hypertarget{references}{%
\section*{References}\label{references}}
\addcontentsline{toc}{section}{References}

\hypertarget{refs}{}
\begin{CSLReferences}{1}{0}
\leavevmode\hypertarget{ref-Sunny}{}%
(2024d). In \emph{GitHub repository}. GitHub.
\url{https://github.com/SunnySuite/Sunny.jl}

\leavevmode\hypertarget{ref-SpinWaveGenie}{}%
(2024a). In \emph{GitHub repository}. GitHub.
\url{https://github.com/SpinWaveGenie/SpinWaveGenie}

\leavevmode\hypertarget{ref-uppasd:2024}{}%
(2024c). In \emph{GitHub}. GitHub.
\url{https://github.com/UppASD/UppASD}

\leavevmode\hypertarget{ref-li:2024}{}%
(2024b). In \emph{GitHub repository}. GitHub.
\url{https://github.com/bingli621/pyLiSW}

\leavevmode\hypertarget{ref-CrystalInfoFramework}{}%
(2024f). In \emph{GitHub repository}. GitHub.
\url{https://github.com/jamesrhester/CrystalInfoFramework.jl}

\leavevmode\hypertarget{ref-Brillouin}{}%
(2024e). In \emph{GitHub repository}. GitHub.
\url{https://github.com/thchr/Brillouin.jl}

\leavevmode\hypertarget{ref-bai:2021}{}%
Bai, X., Zhang, S.-S., Dun, Z., Zhang, H., Huang, Q., Zhou, H., Stone,
M. B., Kolesnikov, A. I., Ye, F., Batista, C. D., \& others. (2021).
Hybridized quadrupolar excitations in the spin-anisotropic frustrated
magnet \(\mathrm{FeI}_2\). \emph{Nature Physics}, \emph{17}(4),
467--472.

\leavevmode\hypertarget{ref-bai:2023}{}%
Bai, X., Zhang, S.-S., Zhang, H., Dun, Z., Phelan, W. A., Garlea, V. O.,
Mourigal, M., \& Batista, C. D. (2023). Instabilities of heavy magnons
in an anisotropic magnet. \emph{Nature Communications}, \emph{14}(1),
4199.

\leavevmode\hypertarget{ref-dahlbom:2024a}{}%
Dahlbom, D. A., Brooks, F. T., Wilson, M. S., Chi, S., Kolesnikov, A.
I., Stone, M. B., Cao, H., Li, Y. W., Barros, K., Mourigal, M., \&
others. (2024). Quantum-to-classical crossover in generalized spin
systems: {T}emperature-dependent spin dynamics of \(\mathrm{FeI}_2\).
\emph{Physical Review B}, \emph{109}(1), 014427.

\leavevmode\hypertarget{ref-dahlbom:2022b}{}%
Dahlbom, D. A., Miles, C., Zhang, H., Batista, C. D., \& Barros, K.
(2022). Langevin dynamics of generalized spins as {SU}({N}) coherent
states. \emph{Physical Review B}, \emph{106}(23), 235154.

\leavevmode\hypertarget{ref-dahlbom:2024b}{}%
Dahlbom, D. A., Thomas, J., Johnston, S., Barros, K., \& Batista, C. D.
(2024). Classical dynamics of the antiferromagnetic {H}eisenberg {S}=1/2
spin ladder. \emph{Physical Review B}, \emph{110}(10), 104403.

\leavevmode\hypertarget{ref-dahlbom:2023}{}%
Dahlbom, D. A., Zhang, H., Laraib, Z., Pajerowski, D. M., Barros, K., \&
Batista, C. D. (2023). Renormalized classical theory of quantum magnets.
\emph{arXiv Preprint arXiv:2304.03874}.

\leavevmode\hypertarget{ref-dahlbom:2022a}{}%
Dahlbom, D. A., Zhang, H., Miles, C., Bai, X., Batista, C. D., \&
Barros, K. (2022). Geometric integration of classical spin dynamics via
a mean-field {S}chr{ö}dinger equation. \emph{Physical Review B},
\emph{106}(5), 054423.

\leavevmode\hypertarget{ref-danisch:2021}{}%
Danisch, S., \& Krumbiegel, J. (2021). Makie.jl: {F}lexible
high-performance data visualization for {J}ulia. \emph{Journal of Open
Source Software}, \emph{6}(65), 3349.

\leavevmode\hypertarget{ref-do:2023}{}%
Do, S.-H., Zhang, H., Dahlbom, D. A., Williams, T. J., Garlea, V. O.,
Hong, T., Jang, T.-H., Cheong, S.-W., Park, J.-H., Barros, K., \&
others. (2023). Understanding temperature-dependent {SU}(3) spin
dynamics in the {S}=1 antiferromagnet {B}a\(_2\){F}e{S}i\(_2\){O}\(_7\).
\emph{Npj Quantum Materials}, \emph{8}(1), 5.

\leavevmode\hypertarget{ref-evans:2014}{}%
Evans, R. F. L., Fan, W. J., Chureemart, P., Ostler, T. A., Ellis, M. O.
A., \& Chantrell, R. W. (2014). Atomistic spin model simulations of
magnetic nanomaterials. \emph{Journal of Physics: Condensed Matter},
\emph{26}(10), 103202.

\leavevmode\hypertarget{ref-hall:1991}{}%
Hall, S. R., Allen, F. H., \& Brown, I. D. (1991). The crystallographic
information file ({CIF}): A new standard archive file for
crystallography. \emph{Foundations of Crystallography}, \emph{47}(6),
655--685.

\leavevmode\hypertarget{ref-kim:2023}{}%
Kim, C., Kim, S., Park, P., Kim, T., Jeong, J., Ohira-Kawamura, S.,
Murai, N., Nakajima, K., Chernyshev, A. L., Mourigal, M., \& others.
(2023). Bond-dependent anisotropy and magnon decay in cobalt-based
{K}itaev triangular antiferromagnet. \emph{Nature Physics},
\emph{19}(11), 1624--1629.

\leavevmode\hypertarget{ref-lane:2024}{}%
Lane, H., Zhang, H., Dahlbom, D., Quinn, S., Somma, R., Mourigal, M.,
Batista, C. D., \& Barros, K. (2024). Kernel polynomial method for
linear spin wave theory. \emph{SciPost Physics}, \emph{17}(5), 145.

\leavevmode\hypertarget{ref-lee:2023}{}%
Lee, M., Schönemann, R., Zhang, H., Dahlbom, D., Jang, T.-H., Do, S.-H.,
Christianson, A. D., Cheong, S.-W., Park, J.-H., Brosha, E., \& others.
(2023). Field-induced spin level crossings within a quasi-{XY}
antiferromagnetic state in {B}a\(_2\){F}e{S}i\(_2\){O}\(_7\).
\emph{Physical Review B}, \emph{107}(14), 144427.

\leavevmode\hypertarget{ref-legros:2021}{}%
Legros, A., Zhang, S.-S., Bai, X., Zhang, H., Dun, Z., Phelan, W. A.,
Batista, C. D., Mourigal, M., \& Armitage, N. (2021). Observation of
4-and 6-magnon bound states in the spin-anisotropic frustrated
antiferromagnet \(\mathrm{FeI}_2\). \emph{Physical Review Letters},
\emph{127}(26), 267201.

\leavevmode\hypertarget{ref-mogensen:2018}{}%
Mogensen, P., \& Riseth, A. (2018). Optim: A mathematical optimization
package for julia. \emph{Journal of Open Source Software}, \emph{3}(24).

\leavevmode\hypertarget{ref-muniz:2014}{}%
Muniz, R. A., Kato, Y., \& Batista, C. D. (2014). Generalized spin-wave
theory: Application to the bilinear--biquadratic model. \emph{Progress
of Theoretical and Experimental Physics}, \emph{2014}(8), 083I01.

\leavevmode\hypertarget{ref-muller:2019}{}%
Müller, G. P., Hoffmann, M., Dißelkamp, C., Schürhoff, D., Mavros, S.,
Sallermann, M., Kiselev, N. S., Jónsson, H., \& Blügel, S. (2019).
Spirit: {M}ultifunctional framework for atomistic spin simulations.
\emph{Physical Review b}, \emph{99}(22), 224414.

\leavevmode\hypertarget{ref-na:2024}{}%
Na, W., Park, P., Oh, S., Kim, J., Scheie, A., Tennant, D. A., Lee, H.
C., Park, J.-G., \& Cheong, H. (2024). Direct observation and analysis
of low-energy magnons with {R}aman spectroscopy in atomically thin
{N}i{PS}\(_3\). \emph{ACS Nano}, \emph{18}(31), 20482--20492.

\leavevmode\hypertarget{ref-nagl:2024}{}%
Nagl, J., Flavián, D., Hayashida, S., Povarov, K. Y., Yan, M., Murai,
N., Ohira-Kawamura, S., Simutis, G., Hicken, T. J., Luetkens, H., \&
others. (2024). Excitation spectrum and spin {H}amiltonian of the
frustrated quantum {I}sing magnet {P}r\(_3\){BWO}\(_9\). \emph{Physical
Review Research}, \emph{6}(2), 023267.

\leavevmode\hypertarget{ref-paddison:2024}{}%
Paddison, J. A. M., Zhang, H., Yan, J., Cliffe, M. J., McGuire, M. A.,
Do, S.-H., Gao, S., Stone, M. B., Dahlbom, D. A., Barros, K., \& others.
(2024). Cubic double perovskites host noncoplanar spin textures.
\emph{Npj Quantum Materials}, \emph{9}(1), 48.

\leavevmode\hypertarget{ref-park:2024c}{}%
Park, P., Cho, W., Kim, C., An, Y., Iida, K., Kajimoto, R., Matin, S.,
Zhang, S.-S., Batista, C. D., \& Park, J.-G. (2024). Contrasting
dynamical properties of single-{Q} and triple-{Q} magnetic orderings in
a triangular lattice antiferromagnet. \emph{arXiv Preprint
arXiv:2410.02180}.

\leavevmode\hypertarget{ref-park:2023a}{}%
Park, P., Cho, W., Kim, C., An, Y., Kang, Y.-G., Avdeev, M., Sibille,
R., Iida, K., Kajimoto, R., Lee, K. H., \& others. (2023). Tetrahedral
triple-{Q} magnetic ordering and large spontaneous {H}all conductivity
in the metallic triangular antiferromagnet {C}o\(_{1/3}\){T}aS\(_2\).
\emph{Nature Communications}, \emph{14}(1), 8346.

\leavevmode\hypertarget{ref-park:2024b}{}%
Park, P., Ghioldi, E. A., May, A. F., Kolopus, J. A., Podlesnyak, A. A.,
Calder, S., Paddison, J. A. M., Trumper, A. E., Manuel, L. O., Batista,
C. D., \& others. (2024). Anomalous continuum scattering and
higher-order van {H}ove singularity in the strongly anisotropic {S}=1/2
triangular lattice antiferromagnet. \emph{Nature Communications},
\emph{15}(1), 7264.

\leavevmode\hypertarget{ref-park:2024a}{}%
Park, P., Sala, G., Pajerowski, D. M., May, A. F., Kolopus, J. A.,
Dahlbom, D. A., Stone, M. B., Halász, G. B., \& Christianson, A. D.
(2024). Quantum and classical spin dynamics across temperature scales in
the \(S\)=1/2 {H}eisenberg antiferromagnet. \emph{Physical Review
Research}, \emph{6}(3), 033184.

\leavevmode\hypertarget{ref-petit:2016}{}%
Petit, S., \& Damay, F. (2016). Spin{W}ave, a software dedicated to spin
wave simulations. \emph{Neutron News}, \emph{27}(4), 27--28.

\leavevmode\hypertarget{ref-rotter:2004}{}%
Rotter, M. (2004). Using {M}c{P}hase to calculate magnetic phase
diagrams of rare earth compounds. \emph{Journal of Magnetism and
Magnetic Materials}, \emph{272}, E481--E482.

\leavevmode\hypertarget{ref-scheie:2023}{}%
Scheie, A., Park, P., Villanova, J. W., Granroth, G. E., Sarkis, C. L.,
Zhang, H., Stone, M. B., Park, J.-G., Okamoto, S., Berlijn, T., \&
others. (2023). Spin wave {H}amiltonian and anomalous scattering in
{N}i{PS}\(_3\). \emph{Physical Review B}, \emph{108}(10), 104402.

\leavevmode\hypertarget{ref-swendsen:1986}{}%
Swendsen, R. H., \& Wang, J.-S. (1986). Replica {M}onte {C}arlo
simulation of spin-glasses. \emph{Physical Review Letters},
\emph{57}(21), 2607.

\leavevmode\hypertarget{ref-togo:2024}{}%
Togo, A., Shinohara, K., \& Tanaka, I. (2024). Spglib: A software
library for crystal symmetry search. \emph{Science and Technology of
Advanced Materials: Methods}, \emph{4}(1), 2384822.

\leavevmode\hypertarget{ref-toth:2015}{}%
Toth, S., \& Lake, B. (2015). Linear spin wave theory for single-{Q}
incommensurate magnetic structures. \emph{Journal of Physics: Condensed
Matter}, \emph{27}(16), 166002.

\leavevmode\hypertarget{ref-wang:2001}{}%
Wang, F., \& Landau, D. P. (2001). Efficient, multiple-range random walk
algorithm to calculate the density of states. \emph{Physical Review
Letters}, \emph{86}(10), 2050.

\leavevmode\hypertarget{ref-watson:2022}{}%
Watson, G. R., Cage, G., Fortney, J., Granroth, G. E., Hughes, H.,
Maier, T., McDonnell, M., Ramirez-Cuesta, A., Smith, R., Yakubov, S., \&
others. (2022). Calvera: A platform for the interpretation and analysis
of neutron scattering data. \emph{Smoky Mountains Computational Sciences
and Engineering Conference}, 137--154.

\leavevmode\hypertarget{ref-zhang_batista:2021}{}%
Zhang, H., \& Batista, C. D. (2021). Classical spin dynamics based on
{SU}({N}) coherent states. \emph{Physical Review B}, \emph{104}(10),
104409.

\leavevmode\hypertarget{ref-zhang:2024}{}%
Zhang, Hao, \& Lin, S.-Z. (2024). Multipolar skyrmion crystals in
non-{K}ramers doublet systems. \emph{Physical Review Letters},
\emph{133}(19), 196702.

\leavevmode\hypertarget{ref-zhang:2023}{}%
Zhang, H., Wang, Z., Dahlbom, D. A., Barros, K., \& Batista, C. D.
(2023). \(\mathrm{CP}^2\) skyrmions and skyrmion crystals in realistic
quantum magnets. \emph{Nature Communications}, \emph{14}(1), 3626.

\end{CSLReferences}

\end{document}